\documentclass[prb,floats,superscriptaddress,showpacs]{revtex4}
\def\@dotsep{4.5}
\makeatother
\usepackage{amsmath}
\usepackage{graphicx}
\newcommand{\funo} {f^{(1)}}    
\begin{document}
\date{\today}
\title{Theory of thermostatted inhomogeneous granular 
fluids: a self-consistent density functional description.}

\author{Umberto Marini-Bettolo-Marconi}
\affiliation{Dipartimento di Fisica, Via Madonna delle Carceri,
68032 Camerino (MC), Italy}
\author{Pedro Tarazona}
\affiliation{Departamento de Fisica Te\'orica de la Materia Condensada
and Instituto Nicol\'as Cabrera,
Universidad Autonoma de Madrid, E-28049 Madrid, Spain}
\author{Fabio Cecconi}
\affiliation{INFM Center for Statistical Mechanics and Complexity,
and Institute for Complex Systems CNR
Via dei Taurini 19, 00182 Rome Italy.}

\begin{abstract}
The authors present a study of the non equilibrium statistical 
properties of a one dimensional hard-rod
fluid dissipating energy via inelastic collisions and
subject to the action of a Gaussian heat bath, simulating
an external driving mechanism.
They show that the description of the fluid based on the one-particle
phase-space reduced distribution function, in principle necessary
because of the presence of velocity dependent collisional dissipation,
can be contracted to a simpler description in configurational space. 
Indeed, by means of a multiple-time scale method the authors derive a 
self-consistent governing
equation for the particle density distribution function.
This equation is similar to the
dynamic density functional equation employed in the study of 
colloids, but contains additional terms taking into 
account the inelastic nature of the fluid.
Such terms cannot
be derived from a Liapunov generating functional and
contribute not only to the relaxational properties,
but also to the non equilibrium steady state properties.
A validation of the theory against molecular dynamics simulations
is presented in a series of cases, and good agreement is found.

\end{abstract}
\maketitle

\section{Introduction}
Granular fluids (GFs) represent one of the current
paradigms of open non-equilibrium systems and, for this reason,  
in the last two decades have been the subject of a
huge amount of experimental, numerical and theoretical 
studies~\cite{Generali1,Generali2,Generali3,Generali4,Generali5,
Generali6,Generali7}.
GFs can be conveniently modeled as assemblies 
of macroscopic particles, 
experiencing instantaneous binary collisions  during which
a fraction of the  kinetic energy is dissipated, i.e., transferred
into internal degrees of freedom.
Under the
action of a vigorous external driving force, GFs 
may appear similar to
ordinary molecular fluids, but crucial differences remain because
inelasticity leads to the appearance of a series of peculiar behaviors,
such as clustering, non-Gaussian velocity distribution, and 
velocity correlations.
These phenomena have no counterparts in molecular fluids 
and render the study of GFs difficult but particularly fascinating.

In spatially uniform systems, 
relations have been obtained between static average quantities such as 
density, kinetic
temperature and pressure, which may be regarded as the
analogue of the equation of state.  In addition, 
a granular hydrodynamics has been developed  
which, due to the inelasticity of collisions, 
differs nontrivially from 
standard hydrodynamics.
The majority of these studies focus on large scale properties of the fluid.
However, in strongly inhomogeneous systems, 
the connection between the
microscale typical of the particles and the macroscale is still
incomplete.  
Recently, some authors~\cite{Mazenko} have proposed 
phenomenological theories, based on local 
mass and momentum conservation laws, incorporating nonideal gas effects via 
an effective free energy functional suitably designed to describe 
the spontaneous formation of loosely and densely packed regions.
This approach sounds very appealing because 
the free energy density functional~\cite{DDFliterature}, besides
being a method computationally simple and physically clear,
has proven to be a useful tool in the theory of
nonuniform fluids with applications to interfacial and freezing phenomena. 
The basic assumption of all density functional theories is that the
thermodynamic potential of a nonuniform system may be approximated 
knowing the structural and thermodynamic properties of the corresponding 
uniform system.
Two questions are in order before proceeding to generalize the density 
functional theory (DFT) to granular materials:
Does the same method offer any new insight in this new area? 
 How far meaningful concepts for standard molecular fluids,  
such as free energy and chemical potential, can 
be extended to systems which are not at thermodynamic equilibrium?
The answer to the second question seems to be desolately
negative and therefore in order to construct a theory
of nonuniform GF, alternative techniques 
not involving free energy functional derivations 
have to be developed.
A step towards this new direction has been 
recently made and an equation of evolution
for the particle density was obtained
in the case of nonequilibrium colloidal systems~\cite{Tarazona1}.
Interestingly, such an equation is very similar to the equation 
considered within
the dynamic density functional method~\cite{Tarazona2}. It  
differs from the latter only for the presence of some higher
order corrections.
The derivation of Ref.~\cite{Tarazona1} employs standard tools of kinetic 
theory and in particular the revised Enskog equation~\cite{Beijeren,Ernst}.
In this paper we generalize
such an approach to fluids of inelastic hard core
particles subject to a stochastic forcing. 

We consider a one dimensional model of granular fluid which is simple
enough as to lend itself to analytic work, but is endowed with a 
sufficient complexity as to display inhomogeneous behavior~\cite{Kadanoff,
Sela,Mcnamara,Mackintosh,
Barrat,Bennaim,Baldassarri,Santos1,CDBP,Natali,Costantini}. 
 One dimensional models may play a useful role 
since they can be employed to test approximations of more general
applicability and allow us to link easily the structural
properties to the dynamical behavior.
Moreover, at equilibrium the qualitative differences between one dimension
and higher
dimensions appear in the development of long-range ordering or any kind
of phase transitions, but not in the short-range packing structure.
In the collision kinetics, the qualitative difference between
one dimension 
and higher dimensions appears in purely inertial and elastic systems
through the conservation of the velocity distributions despite the
collisions. The role of the bath, and also the inelastic collisions,
kills the peculiarity of the one dimensional case.
A basic feature of this work is the assumption of a uniform thermostat
to describe the external energy supply. 
The balance between thermostatting 
and dissipation mechanism 
gives rise to nonequilibrium steady states which are achieved without
fine tuning of the model parameters. Few kinds of thermostats have been
employed in the literature, namely, the white noise thermostat
~\cite{Montanero}, the Gaussian thermostat~\cite{Santos2}
and the Langevin thermostat~\cite{Pagnani,Hayakawa,Biwell} which includes
both the white noise term and the friction force proportional to
the velocity of the particles~\cite{Carrillo}. The present study is based on 
the Langevin thermostat because it can be easily realized in numerical
experiments and lends itself to a great deal of analytical
work.
 
The paper is organized as follows. In Sec. \ref{Model}, we introduce the
equations describing the dynamics of the stochastically
driven inelastic hard-rod fluid model. We start from the stochastic
equations for the trajectories of each particle and discuss how these
can be reduced under some suitable approximations to the
Fokker-Planck-Boltzmann equation for the single-particle phase-space
distribution. Next, in order to make analytical progress, we separate
the velocity and the spatial dependence of the distribution function
and obtain an infinite hierarchy of coupled integro-differential
equations. In Sec.~\ref{Homogeneous} we analyze the 
steady state uniform properties of the system. 
In Sec.~\ref{Evolution} we introduce the evolution equation
for the density of the system, which is obtained in Appendix A by applying
the multiple time scale method, and in 
Sec.~\ref{Applications} we perform a series of 
numerical tests of our theory using a few selected examples.  
The tests compare the two levels of description:
the results obtained studying the single particle trajectories with
those obtained from the density equation. 
Finally, conclusions are drawn in Sec.~\ref{Conclusions}.

\section{Model} \label{Model}
Let us consider a one dimensional fluid consisting of
$N$ identical inelastic hard rods of mass $m$,
length $\sigma$, coefficient of restitution $\alpha$,
positions $x_i$, and velocities $v_i$, with $i=1,...,N$.
For the sake of generality we also consider
an arbitrary external force, $f_{ext}(x)$.    
When the separation of particles $i$ and $i+1$ 
is $\sigma$ a binary inelastic collision occurs.
The collision conserves the total momentum and is described by 
the linear transformation:
\begin{eqnarray}
v_1'&=&v_1+\frac{1+\alpha}{2\alpha}(v_2-v_1)
\\\nonumber
v_2'&=&v_2-\frac{1+\alpha}{2\alpha}(v_2-v_1) \qquad .
\label{col2}
\end{eqnarray}
connecting the precollisional velocities (primed
symbols) and the postcollisional velocities (unprimed symbols).
Since 
in a single collision the amount of total kinetic energy,
\begin{equation}
\Delta E=-\frac{m}{4}(1-\alpha^2)(v_1'-v_2')^2 \;, 
\label{u2}
\end{equation}
is dissipated, without energy injection the particles would come to rest. 
On the contrary, a steady state regime can be reached
if the energy loss through collisions is balanced by an energy injection
that we assume to be realized by the combination of a friction force 
$-m\gamma v_i$ and a stochastic force $\xi_i(t)$. These two forces represent
the Langevin thermostat.
The complete dynamics
can be represented by
the following 2N coupled stochastic differential equations:
\begin{eqnarray}
&\frac{d x_i}{dt} &= v_i \\
& m \frac{d v_i}{dt}& = -m \gamma v_i +f_{ext}(x_i)+f_i^{coll}+\xi_i(t)
\label{kramers}
\end{eqnarray}
where $f_i^{coll}$ indicates symbolically the resultant of the 
impulsive forces acting on particle $i$ in possible 
hard core collisions against other particles.
The stochastic force $\xi_i(t)$
has  zero average $\langle \xi_i(t)\rangle$ and white noise correlation 
\begin{equation}
\langle \xi_i(t)\xi_j(t') \rangle  = 2 \gamma m T_0\delta_{ij} \delta(t-t')\;,
\end{equation} 
The amplitude $T_0$ is the ``heat-bath
temperature'' and $\langle \cdot \rangle$ indicates the average over a
statistical ensemble of realizations of the noise. 
A statistical 
description of the system in terms of the one-particle phase space
distribution $\funo(x,v,t)$, giving the number of particles
in the volume element $(x,x+dx,v,v+dv)$, can be worked out  
by taking the average over all realizations of the stochastic noise
(see. Ref.~\cite{Tarazona1}). 
Moreover,
the distribution $\funo(x,v,t)$ evolves according to the 
governing equation
\begin{eqnarray}
\frac{\partial}{\partial t}\funo(x,v,t)+
\Bigr [v \frac{\partial}{\partial x}+\frac{f_{ext}(x)}{m}
\frac{\partial}{\partial v} \Bigr]\funo (x,v,t)=
\gamma \Bigr[\frac{\partial}{\partial v} v  
+\frac{T_0}{m}\frac{\partial^2}{\partial v^2} \Bigr] \funo(x,v,t)+
k(x,v,t).
\label{fokker}
\end{eqnarray} 
In the left hand side, the term between the square brackets describes the 
free streaming of the particles
subject to the external force $f_e(x)$, the first term 
in the right hand side is the one-particle  
Fokker-Planck collision term representing the interaction with the 
heat bath, while
$k(x,v,t)$ describes the collisions  
among the particles.
We treat these interactions within the
revised Enskog theory (RET), developed by Ernst and van 
Beijeren~\cite{Beijeren}.
The RET for elastic collisions is accurate over the entire fluid range
and describes the crystal phase too. It has been generalized to 
the inelastic regime and used to derive transport 
coefficients~\cite{Brey,Garzo}.
We write the RET collision operator as:
\begin{eqnarray}
&&k(x,v,t)=
\sum_{s=\pm 1} 
\int dv_2   \Theta(s v_{12})(s v_{12})\\\nonumber
&&\times \Bigr[\frac{1}{\alpha^2}g_2(x,x-s\sigma|\rho)
\funo(x,v_1',t)\funo(x-s\sigma,v_2',t)-
g_2(x,x+s\sigma|\rho)\funo(x,v_1,t)\funo(x+s\sigma,v_2,t)\Bigr]
\label{ret2} 
\end{eqnarray} 
Notice that at variance
with the elastic case, a quadratic factor $\alpha^{-2}$
in the gain term, specific to granular gases, appears.
One power is the consequence of the Jacobian 
$dv_1' dv_2'=\frac{1}{\alpha}dv_1 dv_2$ and the second power
stems from the reflection law $v_{12}=-\alpha v_{12}'$.
The sum over $s=\pm 1$  is the analogue in $d=1$ of
the integration over the d-dimensional surface of the hyper-sphere
with radius $\sigma$. 
The RET embodies  
spatial correlations through the hard-rod pair correlation function,
$g_2(x,x\pm\sigma;n)$ evaluated at contact.
As a simplifying approximation, to obtain a theory at the
level of the one-particle distribution, we
take $g_2(x,x\pm\sigma|\rho)$ to be given by its equilibrium 
value~\cite{Percus}
evaluated when the local density is $\rho(x,t)$: 
\begin{equation}
g_2(x \pm \sigma|\rho)=\frac{1}{1-\eta (x \pm\frac{\sigma}{2})}.
\label{g2}
\end{equation}
The time and density dependence occurs entirely via the
local packing fraction
$\eta(x,t)=\int_{x-\sigma/2}^{x+\sigma/2}dx' \rho(x',t)$.
Therefore,  the collision operator Eq.~(\ref{ret2}) is approximated by
an explicit non-local functional of the one-particle density distribution,
with the terms $f^{(1)}(x,v_1,t) f^{(1)}(x\pm \sigma ,v_2,t)$, set by the
collision distance, and the nonlocal density dependence through 
$\eta (x\pm\sigma/2)$, to include the particle correlations.

In the following we shall employ the non dimensional set of variables
which are obtained by measuring the velocities in units of the thermal
velocity  $v_T=\sqrt{k_B T_0/m}$ 
and lengths in unit of $\sigma$, i.e. $V\equiv v/v_T$ and 
$X\equiv x/\sigma$. 
The remaining variables can be non-dimensionalized
according to the transformations
$\tau \equiv t v_T/sigma)$, $\Gamma=\gamma\sigma/v_T$.
$F(X)\equiv \sigma f_{ext}(x)/mv_T^2$.
Finally, the distribution function and the collision term are
rescaled according to the transformations:
$P(X,V,\tau)\equiv \sigma v_T \funo(x,v,t)$ and
$K(X,V,\tau)\equiv \sigma^2 k(x,v,t)$.

Equation (\ref{fokker}) can be cast 
in the following non dimensional form: 

\begin{equation}
\frac{1}{\Gamma}\frac{\partial P(X,V,\tau)}{\partial \tau}
=L_{FP} P(X,V,\tau)
-\frac{1}{\Gamma}V \frac{\partial }{\partial X}  P(X,V,\tau) 
-\frac{1}{\Gamma}F(X,\tau) \frac{\partial }{\partial V}  P(X,V,\tau)+
\frac{1}{\Gamma} K(X,V,\tau)
\label{kramers0} 
\end{equation}
where we have introduced Fokker-Planck operator, $L_{FP}$ by the equation: 
\begin{equation}
L_{FP} P(X,V,\tau) =\frac{\partial}{\partial V}\Bigl[
\frac{\partial }{\partial V }+V\Bigl]  P(X,V,\tau)
\label{fokkerp}
\end{equation}
The eigenfunctions of $L_{FP}$ read explicitly
\begin{equation}
H_{\nu}(V)\equiv \frac{1}{\sqrt{2\pi}}
(-1)^{\nu} \frac{\partial^{\nu}}{\partial V^{\nu}} \exp(-\frac{1}{2}V^2)
\end{equation}
and correspond to discrete eigenvalues $\nu =0,-1,-2,..\;$.
We separate the velocity from the spatial dependence
by expanding, over the basis set $H_{\nu}(V)$,
both the phase-space distribution 
\begin{equation}
P(X,V,\tau)=\sum_\nu \Phi_{\nu}(X,\tau)H_{\nu}(V)
\label{exp1}
\end{equation}
and the collision term
\begin{equation} 
K(X,V,\tau)=\sum_\nu C_{\nu}(X,\tau)H_{\nu}(V).
\label{exp2}
\end{equation}
As shown in Appendix A, the coefficients $C_{\nu}(X,\tau)$ can be expressed 
as nonlocal products
of the moments $\Phi_{\nu}(X,\tau)$
with coefficients which are nonlocal functionals of the density distribution.
Substituting Eqs.(\ref{exp1}) and (\ref{exp2})
in Eq.~(\ref{kramers0}) and using the orthogonality of the 
basis set $H_{\nu}(V)$,  we obtain
a system of coupled equations: 
for the moments $\Phi_{\nu}(X,\tau)$ which can be written in compact
form as:
\begin{eqnarray}
\Bigl[\frac{\partial \Phi_{\nu}(X,\tau)}{\partial \tau}
+\Gamma \nu\Phi_{\nu}(X,\tau)- C_\nu(X,\tau)\Bigl] +
(\nu+1) \frac{\partial \Phi_{\nu+1}(X,\tau)}{\partial X}
+\Bigl[\frac{\partial}{\partial X}-F(X)\Bigl]\Phi_{\nu-1}(X,\tau)=0
\label{brinkmanhiera}
\end{eqnarray}
with $\Phi_{-1}=0$. 

We identify
the moment $\Phi_0(X,\tau)=\int dV P(X,V,\tau)$ with the number density,
$\Phi_1(X,\tau)=\int dV V P(X,V,\tau)$ with the
momentum density and 
$\Phi_2(X,\tau)+\Phi_0(X,\tau)/2 = 1/2\int dV V^2 P(X,V,\tau)$ with the
kinetic energy density.
For $\nu=0,1,2$,  Eq.~(\ref{brinkmanhiera}) 
encodes the balance equations for these moments,
i.e. the hydrodynamic equations characterizing 
a viscous onedimensional fluid~\cite{footnote1}.
Finally, we introduce a local kinetic temperature, often
called granular temperature, via the definition:
\begin{equation} 
T(X,\tau)=(\langle V^2 \rangle - \langle V \rangle^2)  
 = 1 + 2\frac{\Phi_2(X,\tau)}{\Phi_0(X,\tau)}
 - \bigg[\frac{\Phi_1(X,\tau)}{\Phi_0(X,\tau)}\bigg]^2\;.
\label{tempe1}
\end{equation}

\section{Homogeneous steady state properties} \label{Homogeneous}
\begin{figure}[htb]
\includegraphics[clip=true,width=8.0cm, keepaspectratio,angle=0]
{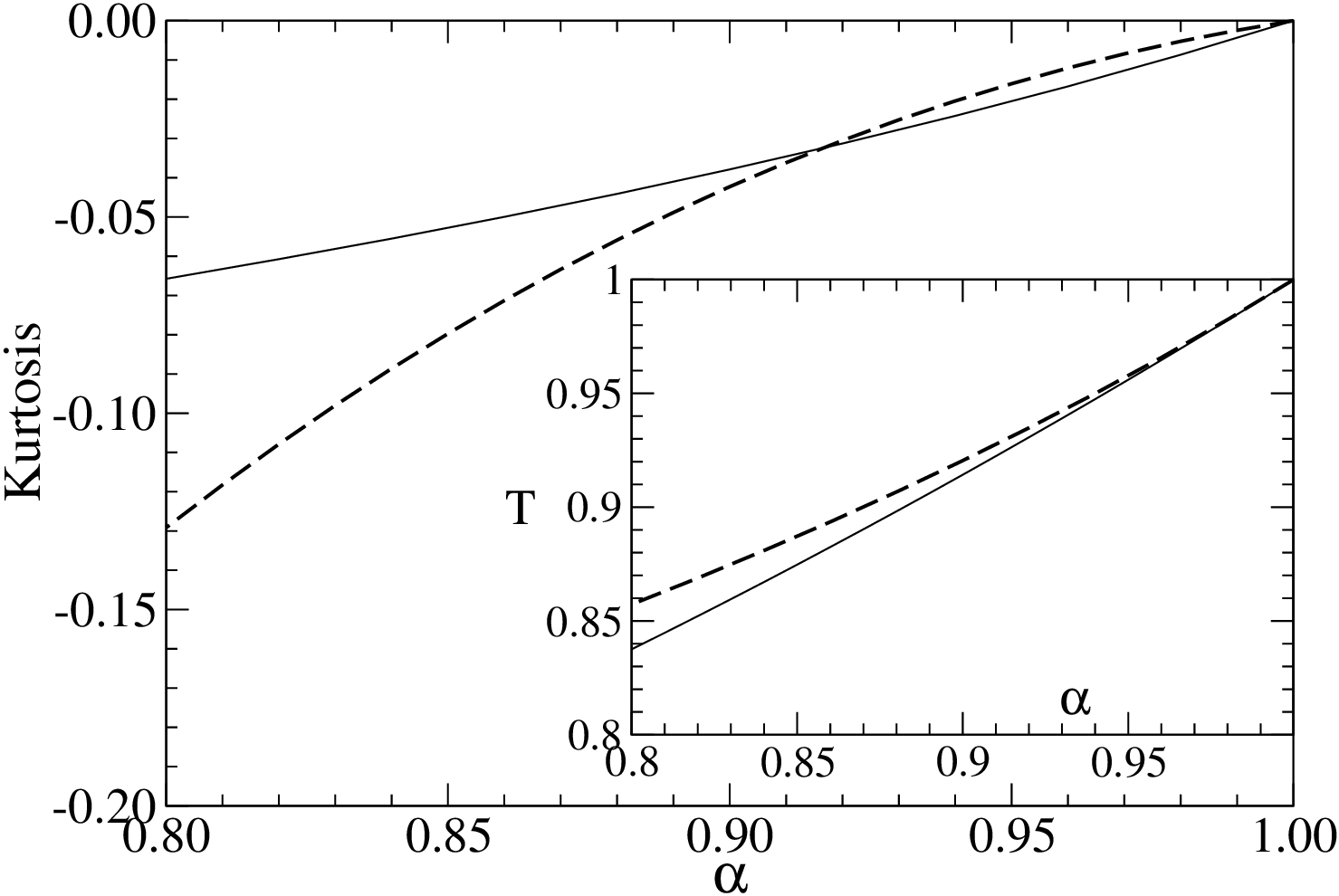}
\caption{Kurtosis of the velocity distribution 
as a function of $\alpha$ for $\rho_0=0.8$ and  $\Gamma=5$.
The dashed line represents the data of the present theory, whilst the
continuous line refers to the results from formula~(\ref{pulvirenti}) 
by Pulvirenti and coworkers.
In the inset we display the corresponding kinetic temperatures.}
\label{fig:kurtosis}
\end{figure}

Before embarking upon the task of solving the evolution  equations, 
we illustrate the peculiarity of the inelastic system by choosing 
the simplest case, namely,   
a time-independent spatially uniform system with $F(X)=0$.
We consider the global velocity distribution 
function $\Psi(V)$, and show that in the steady
state it does not relax to the Maxwellian, as it would occur in the case
of a molecular fluid.
To this purpose let $\Phi_i=\tilde \Phi_i$, where
$\tilde \Phi_i$ are some constants.
The value of the amplitude $\tilde\Phi_2$, through 
Eq.~(\ref{brinkmanhiera}), 
can be expressed in terms of the uniform density $\tilde \Phi_0$:
\begin{equation}
2\Gamma\tilde\Phi_2=C_2=
-\frac{(1-\alpha^2)}{\sqrt{\pi}}g_2
[\tilde\Phi_0^2+3\tilde\Phi_0\tilde\Phi_2+\frac{3}{4}\tilde\Phi_2^2]\;,
\label{br2}
\end{equation}
where the second equality follows from the definition of $C_2$
and the expressions of appendix A.
To first order in the inelasticity parameter $(1-\alpha^2)$ we obtain:
$\tilde\Phi_2= -\epsilon\tilde\Phi_0/2$, where
$\epsilon= [(1-\alpha^2)/(\sqrt{\pi}\Gamma)] g_2 \tilde\Phi_0$.
The procedure can be carried on for values of $\nu$ larger than $2$ 
with the following result:
\begin{equation}
\tilde\Phi_4 = \frac{1}{4\Gamma}C_4\simeq \frac{(1-2\alpha^2)}
{96}\epsilon\tilde\Phi_0
\label{b76}
\end{equation}
\begin{equation}
\tilde\Phi_6= \frac{1}{6\Gamma}C_6\simeq -\frac{(3-12\alpha^2+8\alpha^4)}
{5760}\epsilon\tilde\Phi_0 
\label{b77}
\end{equation}
\begin{equation}
\tilde\Phi_8= \frac{1}{8\Gamma}C_8\simeq -\frac{(15-20\alpha^2+50\alpha^4
-16\alpha^6)}
{215040}\epsilon\tilde\Phi_0 
\label{b78}
\end{equation}
Hence, $\Psi(V)$ can be written as
\begin{eqnarray}
\Psi(V)=&& \frac{e^{-V^2/2}}{\sqrt{2\pi}}
\Bigl[\tilde\Phi_0+(V^2-1)\tilde\Phi_2+(V^4-6V^2+3)\tilde\Phi_4+\\\nonumber
&&(V^6-15V^4+45V^2-15)\tilde\Phi_6+
(V^8-28V^6+210V^4-420V^2+105)\tilde\Phi_8\Bigl]\;.
\label{br3u}
\end{eqnarray}
Two remarks are in order: if we retain only the two leading 
terms in the expansion, the distribution function can be 
approximately rewritten as a Maxwellian,
\begin{equation}
\Psi(V)= \tilde\Phi_0\Bigl[1-(V^2-1)\frac{\epsilon}{2}\Bigl]\frac{e^{-V^2/2}}
{\sqrt{2\pi}} \simeq
\frac{\exp(-\frac{V^2}{2(1-\epsilon)})}{\sqrt{2\pi(1-\epsilon)}}
\tilde\Phi_0\;,
\label{br3b}
\end{equation}
and we interpret $\epsilon$ as the reduced temperature shift induced 
by the inelastic dissipation.
Secondly,
the expansion Eq.~(\ref{br3u}) can be compared with  an exact
solution of Eq.~(\ref{kramers0}),
valid when  $F(X)=0$ and in the limit 
$(1-\alpha)\to 0$~\cite{Pulvirenti}, obtained by
Benedetto et al.. These authors 
showed that a spatially uniform solution,
$\Psi_p(V)$, of Eq.~(\ref{kramers0}) 
is given implicitly by the following nonlinear integral equation:
\begin{equation}
\Psi_p(V)=\frac{e^{-V^2/2}}{\sqrt{2\pi}Z}\exp\Bigl\{-\frac{(1-\alpha) 
g_2}{6\Gamma}
\Bigl[\int_{0}^{\infty}du u^3 \Psi_p(u+V)-\int_{-\infty}^{0}
du u^3 \Psi_p(u+V)\Bigl]\Bigl\}
\label{pulvirenti}
\end{equation}
where $Z$ is the constant which ensures the correct normalization of the 
probability distribution function (PDF).
Interestingly, such a distribution has high-velocity tails which
decay as $\exp(-c|V|^3)$, whereas the central region of the
distribution is approximately a Maxwellian.
Clearly, the high-velocity tails cannot be well reproduced by the 
present expansion, which is applicable when $\Gamma>>1$, but 
the kurtosis associated with Eq.~(\ref{br3u}) compares reasonably
with the kurtosis computed from the distribution $P_p(V)$, as shown in
Fig.~\ref{fig:kurtosis}.

\section{Evolution equation} \label{Evolution}
We shall consider, in the following, the nearly overdamped regime 
$\Gamma>1$ \cite{Grossman,Risken,Gardiner,Vankampen,
Kramers,Smoluchowski,Wilemski}. Since only the particle number is conserved,
one expects that after a transient
of duration of the order of  $\gamma^{-1}$, the momentum 
and the energy current become slaved by the density field. This remark 
allows us to simplify the task posed by
the open hierarchy of Eqs.~(\ref{brinkmanhiera}).
In Ref.~\cite{Tarazona1}, we showed that, for a system
undergoing perfectly elastic collisions, the problem can be treated
conveniently by employing a multiple-time scale technique.
As a result we found a reaction-diffusion 
self-consistent equation
involving only the amplitude $\Phi_0(X,\tau)$.
The evolution of all remaining partial amplitudes $\Phi_{\nu}(X,\tau)$,
($\nu>1$)
could be deduced from the knowledge of $\Phi_0(X,\tau)$. 
Physically, the reason for such a complexity reduction can be attributed
to the fact that 
the marginal velocity probability distribution 
attains its local equilibrium rapidly, in a time span of the order of 
$\gamma^{-1}$, during which the
one-particle density changes slowly.
Indeed, the positional degrees of freedom reach an equilibrium
distribution on a much slower time scale than the velocities.
 
Since the method of solution follows closely the derivation of
Ref.~\cite{Tarazona1}, we report the details of the present case 
in Appendix B and proceed to illustrate the resulting 
equation of evolution. We only recall that the method  
is based on a systematic expansion in powers of $\Gamma^{-1}$ 
which takes into account the fact
that in Eq.~(\ref{kramers0}) the time derivative is multiplied by the 
small parameter $\Gamma^{-1}$. For such a reason a 
multiple time-scale method has to be applied.  
We also need to introduce the following expansions 
of the moments and of the collision integrals in inverse powers of
$\Gamma$:
\begin{equation}
\Phi_{\nu}(X,\tau)= \sum_n \frac{1}{\Gamma^n} \phi_{n\nu}(X,\tau)
\label{a1}
\end{equation}
and
\begin{equation}
C_{\nu}(X,\tau)= \sum_n \frac{1}{\Gamma^n} c_{n \nu}(X,\tau)
\label{a2}
\end{equation}
The key result of the analysis contained
in Appendix B is the following equation for the density 
amplitude $\phi_{00}$:
\begin{equation}
\frac{\partial \phi_{00}}{\partial \tau}(X,\tau) =
\frac{1}{\Gamma} \partial_X \Bigr\{[\partial_X-F(X)]\phi_{00}(X,\tau)
-c_{01}(X,\tau) -
\frac{1}{\Gamma}c_{11}(X,\tau)+\frac{1}{\Gamma} 
\partial_X c_{02}(X,\tau)\Bigr\}\;.
\label{psi0timeb}
\end{equation}
Equation (\ref{psi0timeb}) is the fundamental equation 
of this work and constitutes 
a closed expression, once the collisional 
terms $c_{s\nu}$ and
the amplitudes $\phi_{s\nu}$ are specified in terms
of the scaled density  $\phi_{00}(X,\tau)$.
As we will show below,
the density field $\phi_{00}(X,\tau)$ fully characterizes
the state of the system and slaves the remaining
hydrodynamic fields. Indeed, the amplitudes of the $H_1(V)$ and $H_2(V)$
components are completely determined from the knowledge of
$\phi_{00}(X,\tau)$. Such a complexity reduction occurs
because the density is the only conserved field in our
thermostatted model.
 
Using Eq.~(\ref{colfin}) we obtain at order $\Gamma^{-1}$, the following:
\begin{equation}
c_{01}(X,\tau)= -\frac{(1+\alpha)}{2}\phi_{00}(X,\tau)\Bigl[g_2(X,X+1)
\phi_{00}(X+1,\tau)
-g_2(X,X-1)\phi_{00}(X-1,\tau)\Bigl]
\label{c01}
\end{equation}
and the following at order $\Gamma^{-2}$:
\begin{equation}
c_{02}(X,\tau)= -\frac{(1-\alpha^2)}{2\sqrt{\pi}}\phi_{00}(X,\tau)
\Bigl[g_2(X,X+1)\phi_{00}(X+1,\tau)+g_2(X,X-1)\phi_{00}(X-1,\tau)\Bigl]
\label{c02}
\end{equation}
and
\begin{eqnarray}
c_{11}(X,\tau)&=&\nonumber\\
&&\frac{(1+\alpha)}{\sqrt \pi}\phi_{00}(X,\tau)\Bigl[g_2(X,X+1)
\phi_{11}(X+1,\tau)
+g_2(X,X-1)\phi_{11}(X-1,\tau)\Bigl]\nonumber\\
&-&\frac{(1+\alpha)}{\sqrt \pi} \phi_{11}(X,\tau)\Bigl[g_2(X,X+1)
\phi_{00}(X+1,\tau)
+g_2(X,X-1)\phi_{00}(X-1,\tau)\Bigl]\nonumber\\
&&-\frac{(1+\alpha)}{2}\phi_{00}(X,\tau)\Bigl[g_2(X,X+1)
\phi_{12}(X+1,\tau)-g_2(X,X-1)\phi_{12}(X-1,\tau)\Bigl]\nonumber\\
&&-\frac{(1+\alpha)}{2}\phi_{12}(X,\tau)\Bigl[g_2(X,X+1)
\phi_{00}(X+1,\tau)-g_2(X,X-1)\phi_{00}(X-1,\tau)\Bigl]
\label{c11}
\end{eqnarray}
where
\begin{equation}
\phi_{11}(X,\tau)=-[\partial_X-F(X)]\phi_{00}(X,\tau)+c_{01}(X,\tau)
\label{phi11}
\end{equation}
and 
\begin{equation}
\phi_{12}(X,\tau)=\frac{1}{2}c_{02}(X,\tau).
\label{phi12}
\end{equation}
It is now clear that the quantities $c_{s\nu}(X,\tau)$,
which depend locally on time but nonlocally on space, 
play the role of effective fields because they encode the influence
of the remaining particles on the particle located at $X$.
They are also
functionals of the scaled density  
$\phi_{00}(X,\tau)$, so that Eq.~(\ref{psi0timeb}) is self-consistent and can 
be solved numerically by iteration. 
Relation (\ref{psi0timeb}) is a continuity equation for the particle density,
whose current can be written as
$\Phi_1(X,\tau)= \phi_{11}(X,\tau)/\Gamma+\phi_{21}(X,\tau)/\Gamma^2 $.
 
Interestingly, for $\alpha=1$, Eq.~(\eqref{phi11}) can be recast to
$$
\phi_{11}(X,\tau)=-\phi_{00}(X,\tau)\partial_X\Bigr[
\frac{\delta F^{rod}[\phi_{00}]}{\delta\phi_{00}(X,\tau)}+V_{ext}(X)\Bigr],
$$
where $F^{rod}[\phi_{00}]$ is the hard rod density functional of the 
instantaneous density $\phi_{00}(X,\tau)$. 

The term $c_{02}$ vanishes in the limit $\alpha\to 0$
and describes a tendency of the particles to form denser aggregates
due to their inelasticity.

Also notice that at the zero order in $\Gamma^{-1}$, Eq.~\eqref{psi0timeb} 
may be rewritten as a dynamic density functional (DDF) 
equation~\cite{Tarazona1}, since the
only change with respect to the $\alpha=1$ case \cite{Tarazona2}
is the presence of a prefactor $(1+\alpha)/2$
in the $c_{01}$ term in Eq.~\eqref{c01}.
Hence, if we define  an inelastic free energy density functional as
$F_{\alpha}[\rho]=F_{ideal}[\rho]+ (1+\alpha)/2 F_{excess}[\rho]$
scaling the exact hard-rods excess of the equilibrium case, we would cast
Eq.~\eqref{psi0timeb} into a DDF equation for arbitrary 
values of $\alpha$. 
A particular result would be that, always at the leading order in 
$\Gamma^{-1}$, the equilibrium density profiles 
should be given by the minimum of such inelastic free energy density 
functional.  The results in 
Figs.~\ref{fig:wall} and~\ref{fig:periodic} are qualitatively
consistent with that effect since the reduction of the excess free energy
reduces the oscillations. 
As we shall see below the equation of state for the uniform fluid 
(Eq.~\eqref{col7}) predicts a lowering of the pressure with respect to the 
elastic case  also consistent with such a scaling 
of the nonideal part of the free energy.

Hereafter, we briefly derive some useful relations between the density profile,
the temperature and the pressure in the non uniform steady state where 
the current $\Phi_1$ vanishes.

\subsection{Steady state temperature profile} 
In the limit $\tau\to\infty$
we determine the granular temperature profile,
using the previous results and Eq.~(\ref{tempe1}):
\begin{equation}
T(X)=1+\frac{2}{\Gamma} \frac{\phi_{12}(X)}{\phi_{00}(X)}=
1-\frac{(1-\alpha^2)}{2\Gamma\sqrt \pi}
\Bigl[g_2(X,X+1)\phi_{00}(X+1,\tau)+g_2(X,X-1)\phi_{00}(X-1)\Bigl]
\label{temprofil}
\end{equation}
where we suppressed the time argument and the functions of the
single spatial argument
have to be understood as their asymptotic limiting values
when $\tau\to\infty$. 
The constant $1$ in the {\em r.h.s.} represents
(in our reduced units) the heat-bath temperature, whereas the second term 
is the shift in the local temperature induced by collisions. 
In fact, it amounts to 
the product of three factors: the kinetic energy dissipated, 
the collision rate $\omega_E$ 
(Enskog collision frequency~\cite{Enskog}) and the typical time 
$\gamma^{-1}$ of the heat-bath. The average Enskog frequency 
at each side 
of the particle located at $X$ is:
\begin{equation}
\frac{\omega_E(X\pm 1)}{\gamma}
=\frac{2}{\Gamma\sqrt \pi}g_2(X,X\pm1)\phi_{00}(X\pm 1)
\label{enskog}
\end{equation}
and in the case of a uniform system it reduces to the bulk
Enskog frequency 
$\omega_E=2 v_T\rho\sigma g_2/{\sqrt \pi}$.
where $v_T$ is the thermal velocity of the gas.

\subsection{Steady state pressure profile.}
We now turn our attention to the pressure profile $\Pi(X,\tau)$, which can be 
separated into a kinetic and a collisional contribution
\begin{equation}
\Pi(X)=\Pi_{kin}(X)+\Pi_{coll}(X)\;.
\label{presprofil}
\end{equation}
The total pressure $\Pi(X,\tau)$ is implicitly 
determined  from the momentum balance equation, obtained by considering 
Eq.~(\ref{brinkmanhiera}) with $\nu=1$ in the $\tau\to\infty$ limit 
\begin{equation}
F(X)\Phi_0(X)
-\frac{\partial [\Pi_{kin}(X)+\Pi_{coll}(X)] }{\partial X}=0\;.
\label{hiera1bb}
\end{equation}
We identify  the first term as
\begin{equation}
\Pi_{kin}(X)=\int dV V^2 P(X,V)=
\Phi_{0}(X)+ 2 \Phi_{2}(X)
\label{presid}
\end{equation}
or using the results of the $\Gamma$ expansion, we rewrite
\begin{eqnarray}
\Pi_{kin}(X)=
\phi_{00}(X)+\frac{2}{\Gamma} \phi_{12}(X)\;,
=T(X)\phi_{00}(X)
\label{preskinprofil}
\end{eqnarray}
where we used Eq.~(\ref{temprofil}) to obtain the last equality. 
In Ref.~\onlinecite{Tarazona1} we showed that the
spatial derivative of the collisional pressure 
is related to  the collision integral  via the relation
\begin{equation}
\partial_X\Pi_{coll}(X)=-C_1(X)=-c_{01}(X)
-\frac{1}{\Gamma}c_{11}(X)\;.
\label{prescol}
\end{equation}
By manipulating expressions~(\ref{c01}) and~(\ref{c11})
(see Ref.~\onlinecite{Tarazona1} for details)
we formally integrate Eq.~(\ref{prescol}) with the following result:
\begin{eqnarray}
&&\Pi_{coll}(X)=\frac{(1+\alpha)}{2}
\int_0^1 dz g_2(X-(1-z),X+z)\times\\\nonumber
&&\Bigl\{\phi_{00}(X-(1-z))\phi_{00}(X+z)\\\nonumber
&&+\frac{1}{\Gamma}\Bigl[\phi_{00}(X-(1-z))
\phi_{12}(X+z)+\phi_{12}(X-(1-z)) \phi_{00}(X+z)\Bigl]
\\\nonumber
&&-\frac{2}{\Gamma\sqrt \pi}\Bigl[
\phi_{00}(X-(1-z))\phi_{11}(X+z)-\phi_{11}(X-(1-z))
\phi_{00}(X+z)\Bigl]\Bigl\}\;.
\label{col5}
\end{eqnarray}
In the case of constant density, the z-integration can be trivially 
performed, and we obtain
\begin{equation}
\Pi_{coll}=\frac{(1+\alpha)}{2}\frac{\phi_{00}^2}{1-\phi_{00}}
\Bigl(1+\frac{2}{\Gamma}\frac{\phi_{12}}{\phi_{00}}\Bigl)\;.
\label{col6}
\end{equation}
Finally, by using Eq.~(\ref{preskinprofil})
we cast the equation of state  in the uniform
non equilibrium steady state in the form:
\begin{equation}
\Pi=T\phi_{00}
\Bigl[1+\frac{(1+\alpha)}{2}\frac{\phi_{00}}{1-\phi_{00}}\Bigl]\;.
\label{col7}
\end{equation}
Expressing the temperature as a function
of the density (from Eq. (\ref{temprofil})) 
\begin{equation}
T=1-\frac{(1-\alpha^2)}{\Gamma\sqrt \pi} g_2\phi_{00}\;,
\label{bulktemp}
\end{equation}
we see that Eq.~(\ref{col7}) describes the lowering of
the pressure due to the collisional reduction of the temperature and 
becomes the familiar hard-rod pressure equation for $\alpha=1$.

\section{Applications} \label{Applications}

\subsection{Temporal decay of a small density modulation}
\begin{figure}[htb]
\includegraphics[clip=true,width=8.0cm, keepaspectratio]
{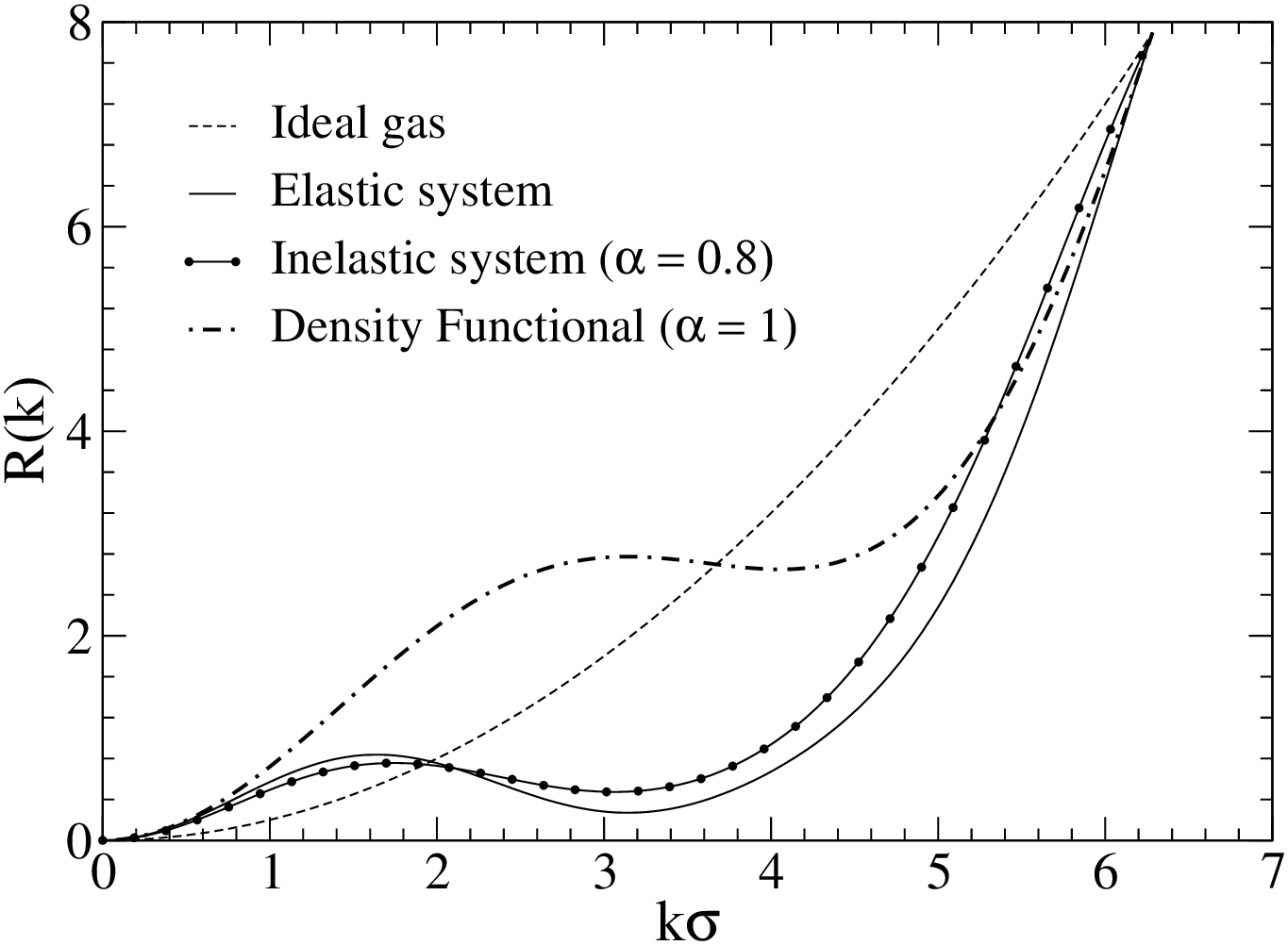}
\caption{Decay rate, $R(K)$ of a small sinusoidal density
perturbation around a constant background $\rho_0=0.68$ as a function of the
the reduced wave-vector $K$. 
The dimensionless friction constant is  $\Gamma=5$.
The non monotonic dependence of $R(K)$ on $K$ increases as the density
increases. Therefore collisions can accelerate or slow down the relaxation
with respect to an ideal gas behavior.  
The dashed line represents the decay rate of a system of non interacting
particles, the dash-dot line the Density functional result, the full line
the system with$\alpha=1$, and the dot-line the inelastic system with   
$\alpha=0.8$. }
\label{fig:due}
\end{figure}

We begin by considering the decay of an infinitesimal
sinusoidal perturbation of wave-vector $K$ with respect to a uniform 
density profile 
and how the relaxation time varies as a function of $K$.
We assume that the sinusoidal density perturbation is small
with respect to some uniform background density, $\Phi_0$ and write:
\begin{equation}
\phi_{00}(X,\tau) = \Phi_0 + \hat\rho_K(\tau)\sin(KX).
\label{uno}
\end{equation}
After some simple algebra we arrive at the following 
equation of evolution for the modulation:
\begin{eqnarray}
\frac{\partial \rho_K(\tau)}{\partial \tau}&=&- R(K)\rho_K(\tau)\nonumber\\
&=&-\frac{K^2}{\Gamma}\rho_K(\tau)\Bigl\{ \Bigl[ 1
+\frac{(1+\alpha)}{2}\Bigl( 2 p_0\sigma\frac{\sin(K)}{K}+ 
\frac{4 (p_0\sigma)^2}{K^2} \sin^2(K/2)\Bigl)\Bigl]\nonumber\\
&&\Bigl[1-2\frac{(1+\alpha)}{\Gamma\sqrt \pi}p_0\sigma 
(1-\cos(K))\Bigl]
\nonumber\\
&-&\frac{(1-\alpha^2)}{\Gamma\sqrt \pi}
\frac{(1+\alpha)}{2}
p_0\sigma
\frac{1}{2}
\Bigl[2 p_0\sigma\frac{\sin(K)}{K}+
\frac{4(p_0\sigma)^2 }{K^2} \sin^2(K/2)\Bigl]\nonumber\\
&-&\frac{(1-\alpha^2)}{\Gamma\sqrt{\pi}}
\Bigl[p_0\sigma(1+\cos(K))+ 
(p_0\sigma)^2\frac{\sin(K)}{K}\Bigl]
\Bigl[1+\frac{(1+\alpha)}{2}p_0\sigma\frac{\sin(K)}{K}\Bigl]
\Bigl\}
\end{eqnarray}
with $p_0 = \Phi_{0}/(1-\Phi_{0})$ being the bulk hard rod reduced pressure.
Taking the long wavelength limit $K\to 0$, we obtain the following expression
for the diffusion coefficient: 
\begin{equation}
D=\lim_{K\to 0} \frac{R(K)}{K^2}=
\frac{1}{\Gamma} \Bigl\{1
+(1+\alpha)(p_0\sigma+ p_0^2\sigma^2/2)
-\frac{(1-\alpha^2)}{\Gamma\sqrt \pi}
\Bigl(2+3 \frac{1+\alpha}{2}\: p_0\sigma\Bigl)
\Bigl(p_0+ p_0^2\sigma/2\Bigl)
\Bigl\}.
\label{alpha}
\end{equation}
As shown in Fig.~\ref{fig:due}, the relaxation time of the inelastic fluid, 
for small values of $K$, is longer than the
corresponding quantity in the elastic fluid. This occurs because, 
being the granular temperature of the former lower, the diffusion
is weaker.
In addition, while the excluded volume favors diffusion
with respect to the non interacting case, the 
inelasticity operates in the opposite direction.
For $K\to 0$ the
temperature field $T(X,\tau)$ is anti-correlated with
the density fluctuation  $\rho(X,\tau)$, and the local maxima of $T$
correspond to the regions where the density is lower. 
On the other hand, for $K$ larger and close to $K=\pi$, 
the temperature maxima
occur in correspondence with the density maxima, thus we observe that
the relaxation time of the inelastic system is shorter than the relaxation
time for $\alpha=1$. 

\subsection{Numerical tests}
\begin{figure}[htb]
\includegraphics[clip=true,width=8.cm, keepaspectratio,angle=0]
{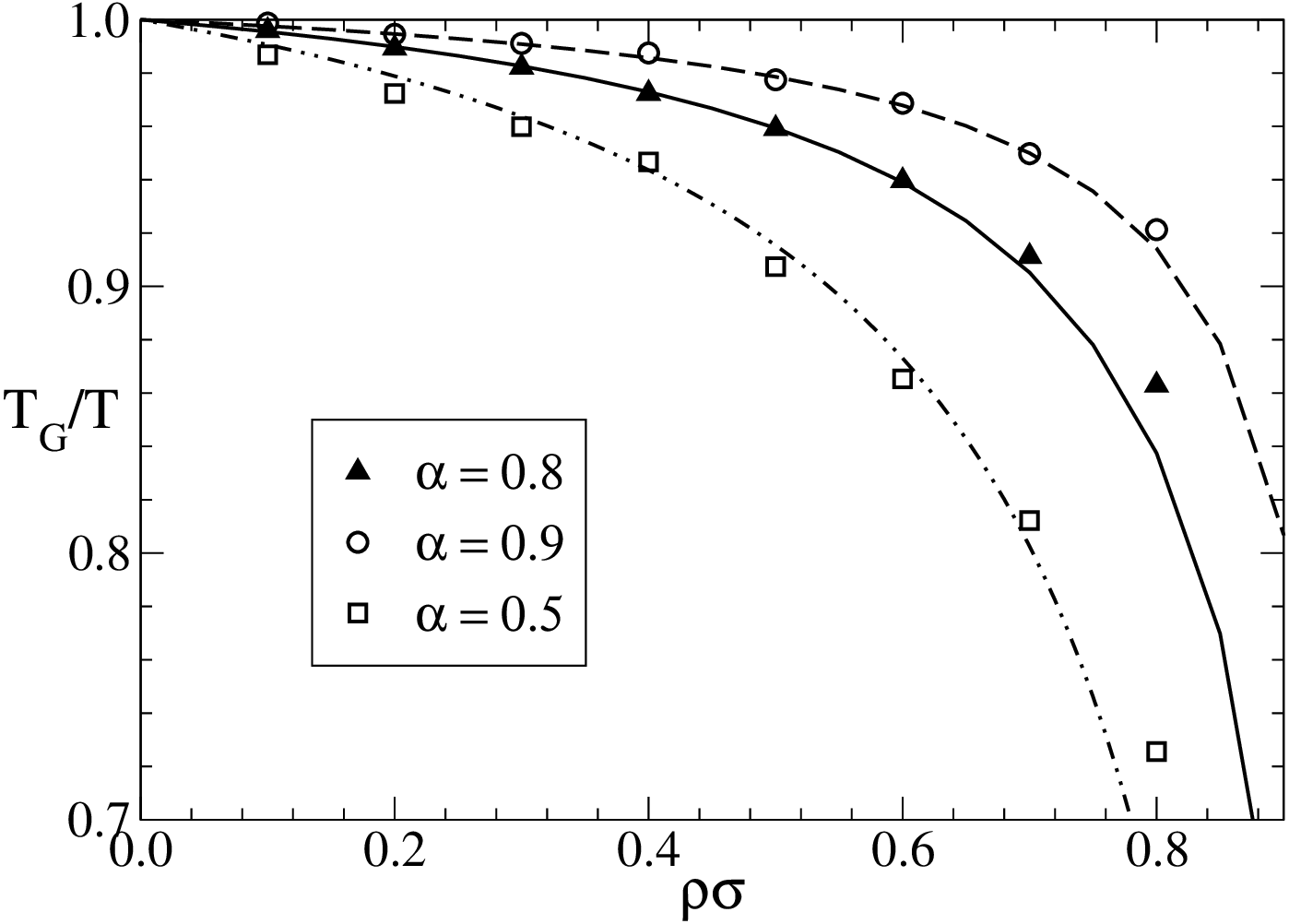}
\caption{Uniform system:
temperature versus reduced density for $\alpha=0.9$, $\alpha=0.8$ and $\Gamma=5$. 
Comparison between Brownian dynamics simulations (points) and 
the predictions of our theory (lines).}
\label{fig:temperatura}
\end{figure}

\begin{figure}[htb]
\includegraphics[clip=true,width=8.cm, keepaspectratio,angle=0]
{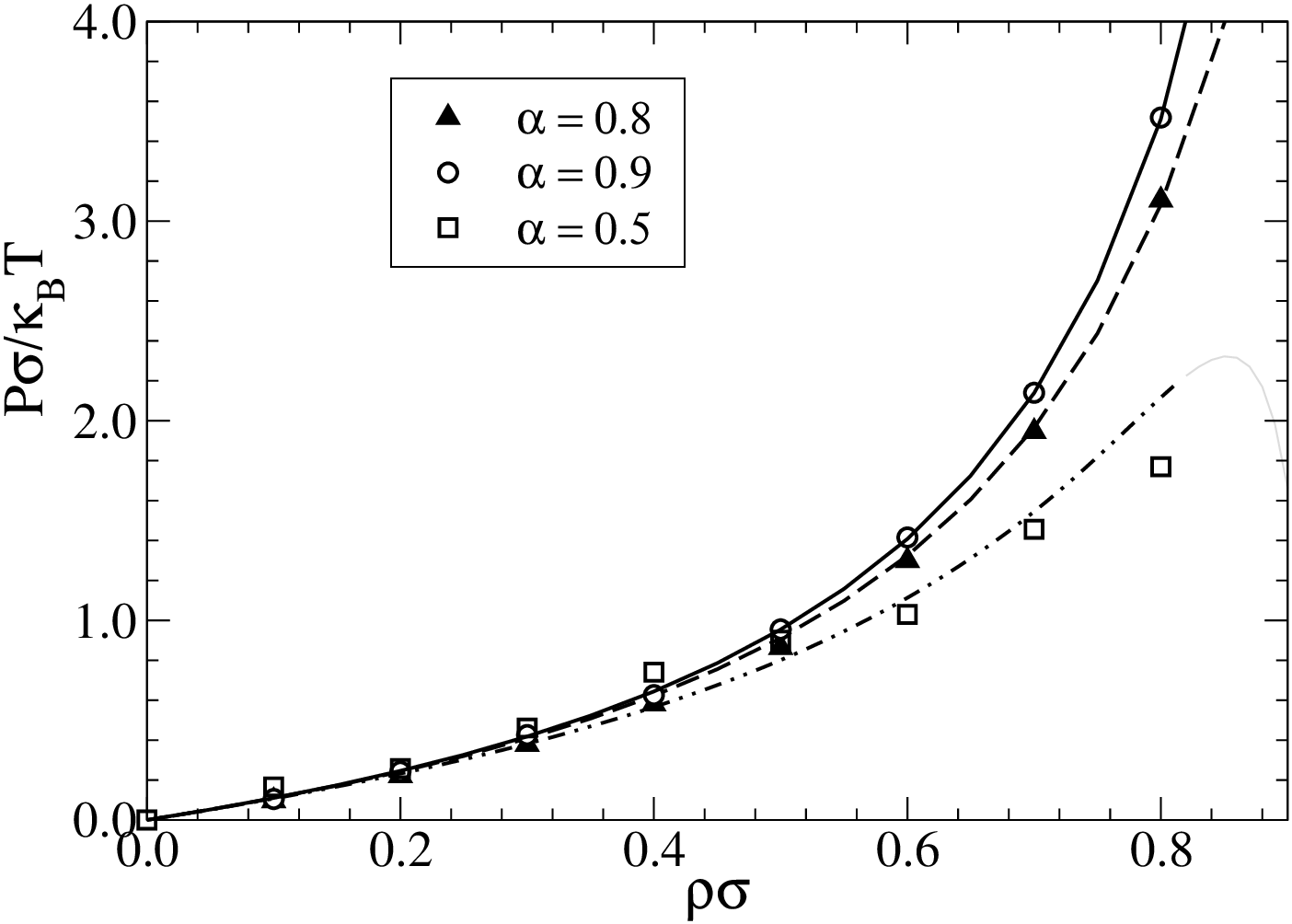}
\caption{Uniform system: pressure versus reduced density 
for $\alpha=0.9$, $\alpha=0.9$ and $\Gamma=5$. Comparison between Brownian 
dynamics simulations (points) and the predictions of our theory (lines).}
\label{fig:pressure}
\end{figure}

To validate the theory we shall compare the  predictions of 
Eq.~(\ref{psi0timeb}) with those obtained by a numerical solution of the
dynamical equations for the trajectories of the particles,
using the algorithm illustrated in Ref.~\onlinecite{CDBP}.
The comparison is performed by considering
an ensemble of ``noise'' histories (typically $10^4$) 
and averaging the observables over such an ensemble. 
The first category of checks concerns the homogeneous static properties 
of the system, namely temperature and pressure.
The  dependence of the temperature on the
density predicted by Eq.~(\ref{bulktemp})
is shown in Fig.~\ref{fig:temperatura}, where it is displayed against
the numerical results obtained at
two different values of the coefficient of restitution.
A similar comparison between pressure [Eq.~(\ref{col7})] and the simulation
results are reported in Fig.~\ref{fig:pressure}, showing a satisfactory 
agreement.
The second category concerns the inhomogeneous static properties, which can
be probed  by  measuring the  response of the model to
some specific non-uniform external perturbations. 

The third category of tests aims, instead, to probe some genuinely
time-dependent properties of the system and we have chosen as
examples the free expansion of a packet of particles initially localized
in a narrow region and
the escape of a packet from a potential well.  

\subsection{Inhomogeneous Steady State Properties}
\begin{figure}[htb]
\includegraphics[clip=true,width=8.cm,keepaspectratio,angle=0]
{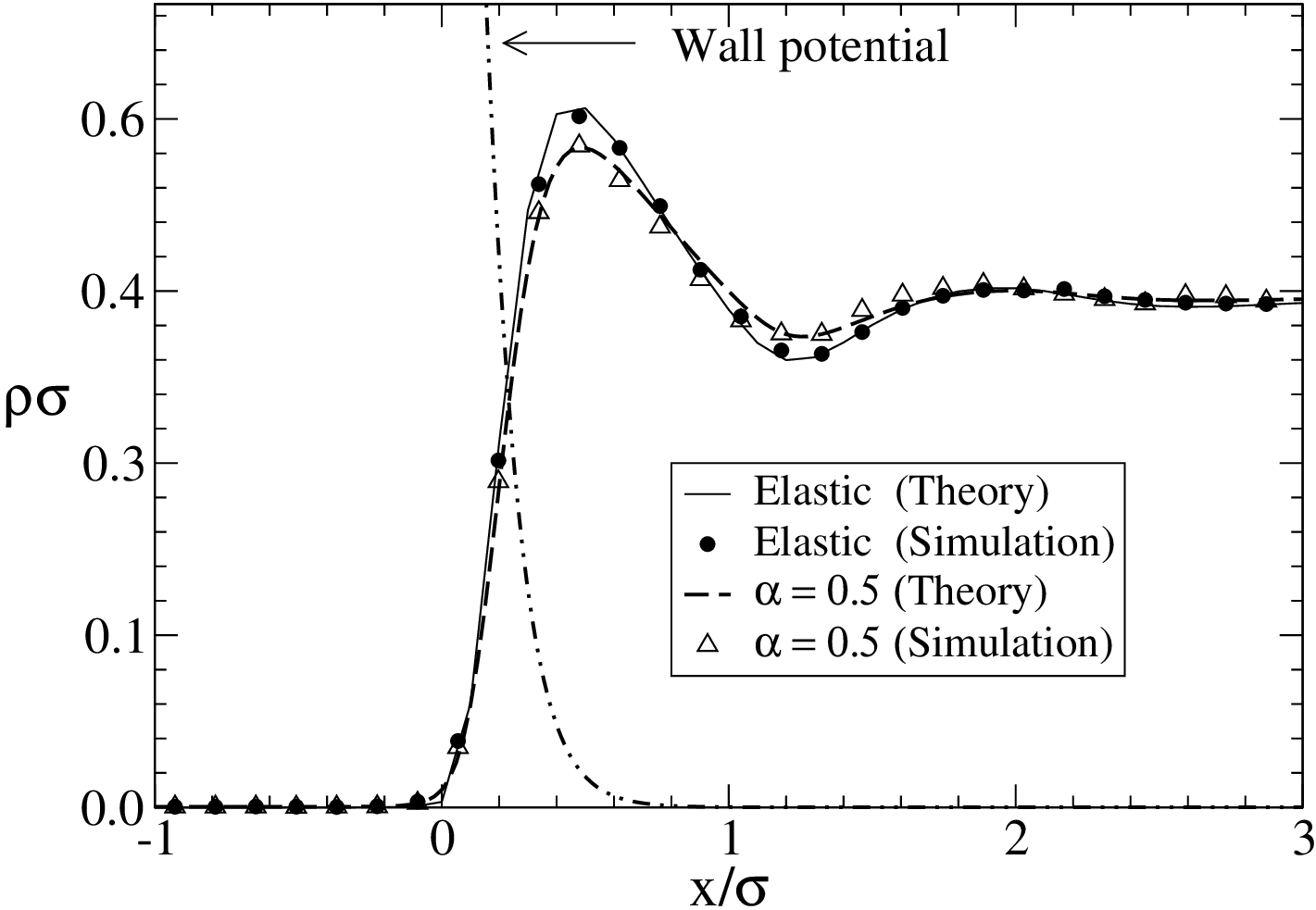}
\caption{Density profiles in the presence
of a repulsive soft wall at $x=0$, 
indicated by a dashed line, and implemented
in the simulations through the potential 
$V(X)=V_0\tanh(X/\xi)$ with $V_0=4$. 
Data refer to $\alpha=1$, $\alpha=0.5$ and $\Gamma=5$. 
Points indicate the results of the simulations whilst the 
lines are the corresponding results from our theory. The 
agreement between simulation an theory is excellent and  
it is also interesting to note that the elastic system  
($\alpha=1$) presents a higher peak near the wall corresponding 
to a larger pressure exerted, toward the wall, on each particle 
by the rest of the 
system.}
\label{fig:wall}
\end{figure}

{\it Soft repulsive potential.}
We consider, first, a fixed external potential of the form
\begin{equation}
V(X)=V_0\tanh(X/\xi),
\label{softwall}
\end{equation}
representing a soft repulsive wall  located at $X=0$ 
and characterized by a softness parameter $\xi=0.2$ and 
height $V_0=4$.
In Fig.~\ref{fig:wall}
we compare the density profiles, obtained from
the stationary solution of the dynamical Eq.~(\ref{psi0timeb}), for two 
values of coefficient of restitution and for $\Gamma=5$, with
the corresponding profiles extracted from molecular dynamics simulation.
The wall perturbs the fluid by inducing a non monotonic profile and a
 stationary state is achieved when hydrostatic equilibrium is reached.

\begin{figure}[htb]
\includegraphics[clip=true,width=8.cm, keepaspectratio,angle=0]
{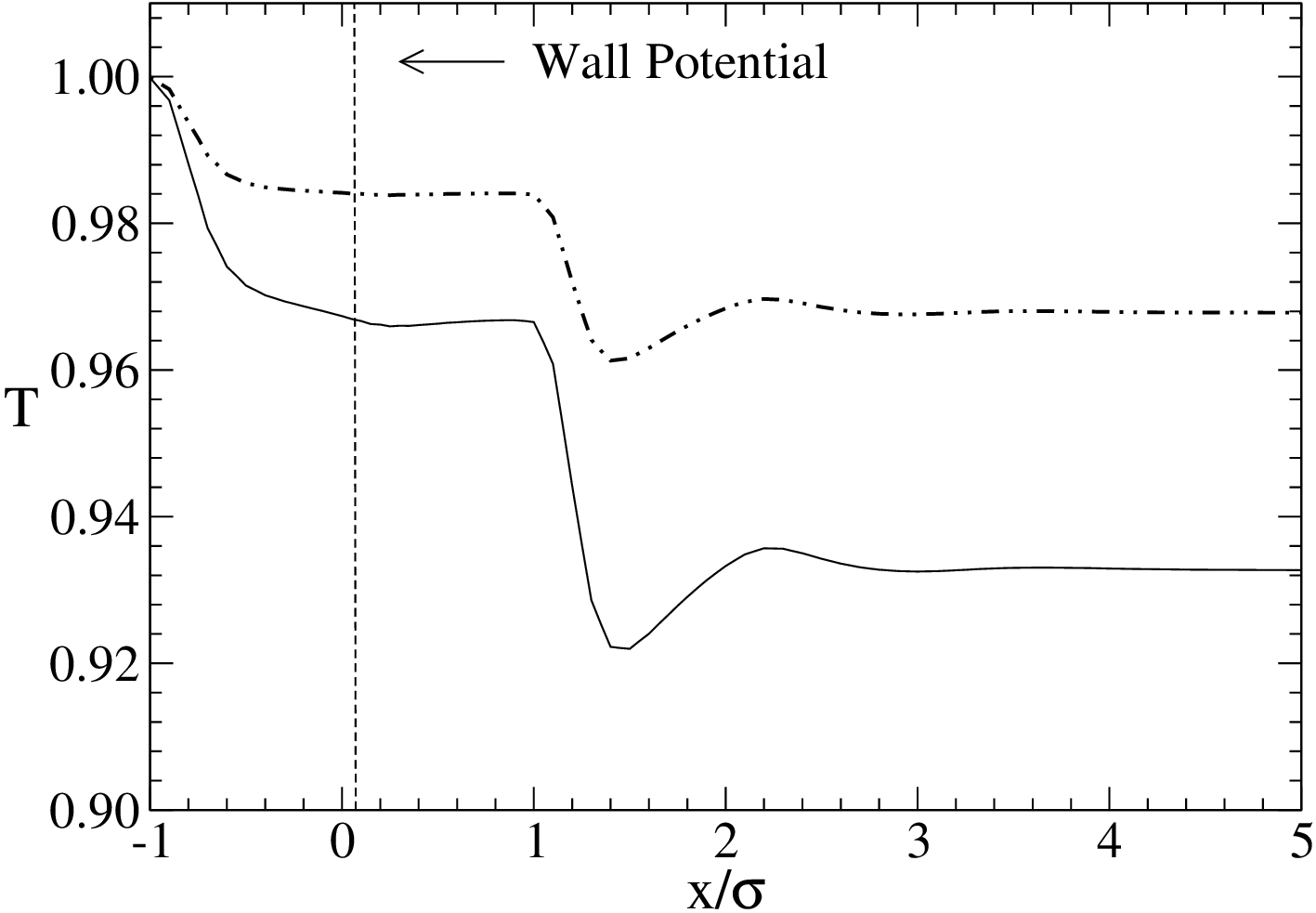}
\caption{Temperature profiles in the presence
of the same repulsive soft wall of Fig.~\ref{fig:wall} 
for a system with $\Gamma=5$ and inelasticity $\alpha=0.8$, $\alpha=0.5$.}
\label{fig:cinquebis}
\end{figure}

Both the MD and the integral equation reveal the same feature:
near the wall the elastic density profile 
is slightly higher than the corresponding profile with $\alpha=0.8$.  
Intuitively such a difference
can be understood by considering that the bulk pressures,
approximately proportional to the corresponding values of the
the density near the wall, display a similar difference. 
The temperature profile, shown in Fig.~\ref{fig:cinquebis},
varies non monotonically
from the value inside the wall to a lower value in the bulk
and is the signature of the non 
equilibrium nature of the system. 

\begin{figure}[htb]
\includegraphics[clip=true,width=8.cm, keepaspectratio,angle=0]
{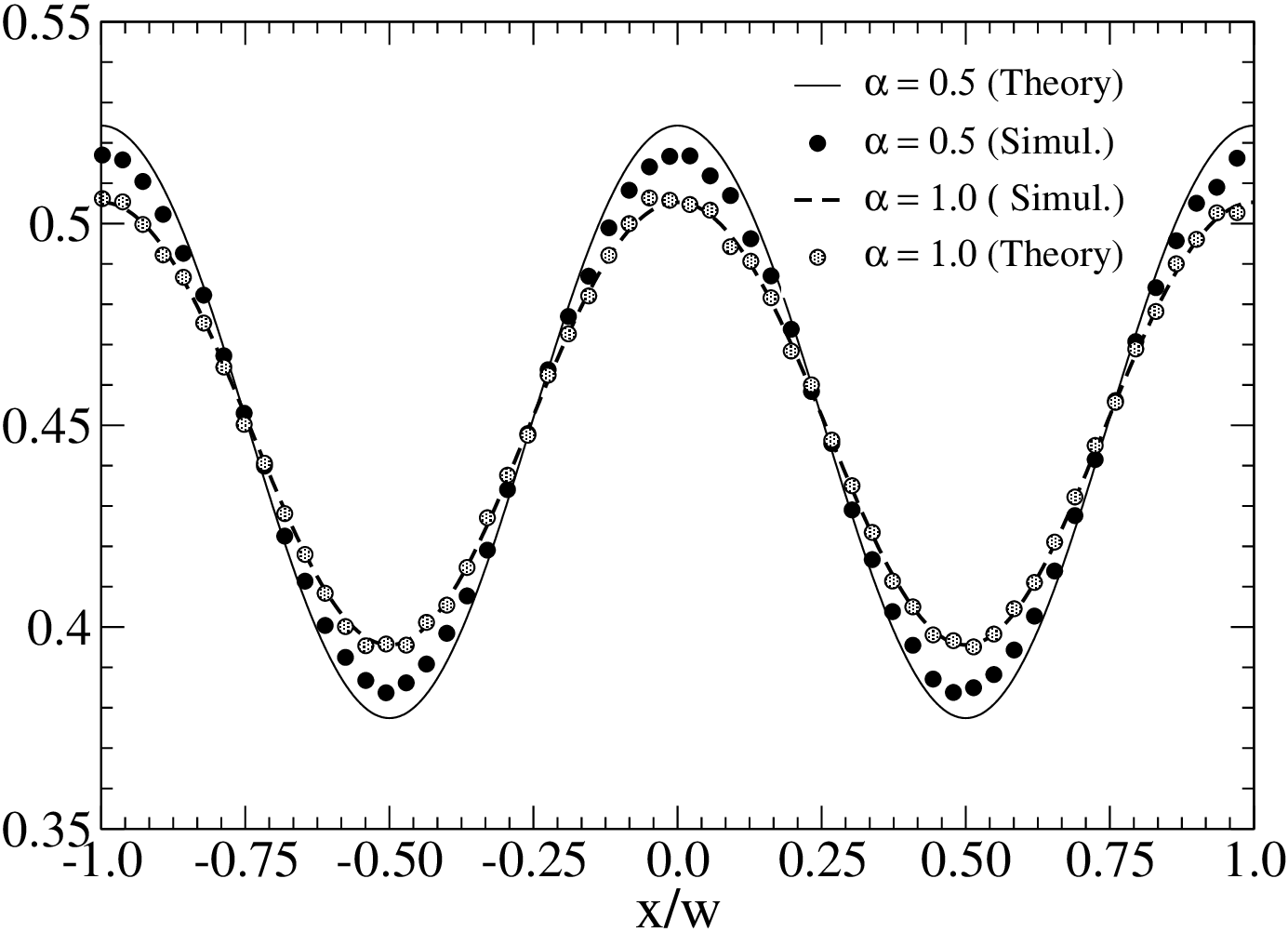}
\caption{Density profiles of a system with average density $\rho = 0.45$ in an 
external potential defined by Eq.~(\ref{periodicpotential}) 
with $w=8$ and $\Gamma=5$.
Black circles correspond to simulation results with $\alpha=0.8$, while 
shaded circles refer to simulations with $\alpha=1.0$. Dashed and full lines 
indicate the corresponding results from the present theory for $\alpha=1.0$ and 
$\alpha = 0.8$ respectively.}
\label{fig:periodic}
\end{figure}

We study, now, the stationary profile induced by a 
static periodic external potential of the form
\begin{equation}
V(X)=V_0 \cos(\frac{2\pi}{w} X).
\label{periodicpotential}
\end{equation} 
For moderate values of the bulk packing fraction
the asymptotic value of the induced density profile turns out to be
modulated with the same period as the potential. 
The height of the peaks
is lower than the corresponding height of the non-interacting case, 
because the hard-core repulsion tends to smear the particles over the wells
away from the minimum energy configuration. On the other hand, 
one can appreciate a difference between the elastic 
and the inelastic case. The latter displays peaks slightly higher and narrower
as shown in Fig.~\ref{fig:periodic},
a fingerprint of the tendency toward clustering induced by the inelasticity
of collisions.


\subsection{Inhomogeneous Dynamical Properties}
\begin{figure}[htb]
\includegraphics[clip=true,width=8.cm, keepaspectratio,angle=0]
{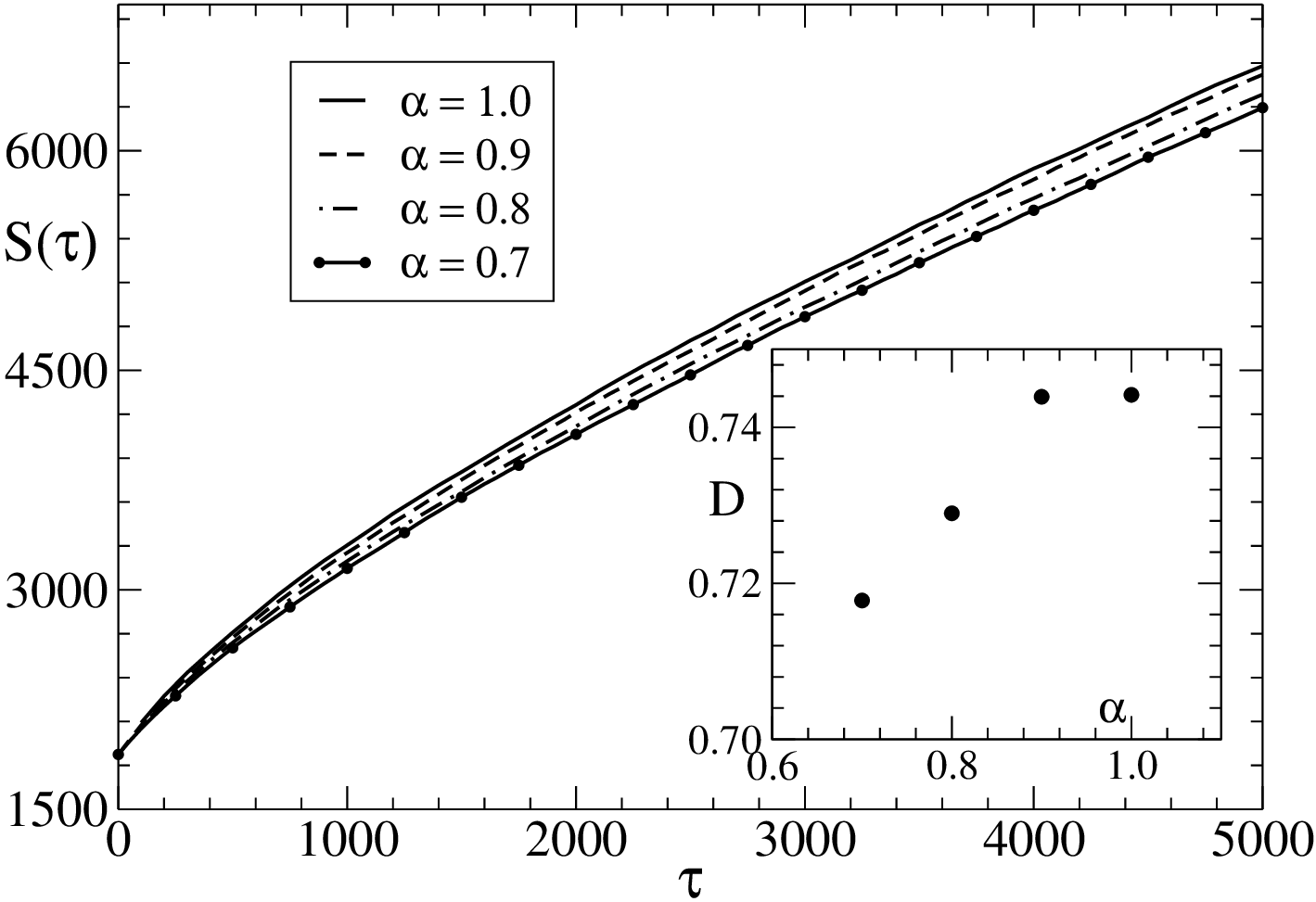}
\caption{Free expansion of a cluster of $128$ particles initially
concentrated over a region of size $150 \sigma$. 
The curves represent the time growth of the variance of the particle 
distribution 
with respect to their center of mass for inelastic systems with 
different values of
$\alpha$ but same $\Gamma=5$. The inset shows the values of the 
diffusion coefficient estimated by the asymptotic slope of the curves.}
\label{fig:diffus}
\end{figure}

We consider the free expansion of $N$ hard-rods in the absence
of external fields. In Fig.~\ref{fig:diffus}, 
we display the evolution of the variance
of the positions of the particles, with respect to their center of mass, 
$W = \frac{1}{N}\sum_i \langle (x_i-x_{cm})^2 \rangle $ for different values
of the inelasticity $\alpha$. The average is meant over different
and independent noise realizations.
In the case of free particles $D=\Gamma^{-1}$, 
we observe a linear growth of this quantity, which is well described by
the diffusive law, $W(\tau) = 2 D\tau$.
In agreement with our analytical prediction of Sec.~VI, the
coefficient $D$ is a growing function of the coefficient of restitution 
and of the number of particles. The first effect can be interpreted
by noticing that smaller values of $\alpha$ correspond to larger
dissipation and thus to lower local kinetic temperatures. Since one expects
$D$ to be proportional to the ratio between temperature and
friction coefficient a lowering of the kinetic temperature determines
a decrease of the spreading. On the other hand, one can compare the
spreading of the same initial configuration in the case of non-interacting
particles. Figure~\ref{fig:diffus} shows that the ideal gas case 
corresponds to a diffusion slower of all cases where the hard core repulsion
is at work. 
Finally, the importance of the corrections to the DDF equation, can be 
appreciated by rescaling the data corresponding to different values
of $\Gamma$  according to
the formula $W/\Gamma$ .
The  free particle case, of course gives a perfect 
collapse, whereas the interacting cases display increasing deviations
as $\Gamma$ decreases. 
\begin{figure}[htb]
\includegraphics[clip=true,width=8.cm, keepaspectratio,angle=0]
{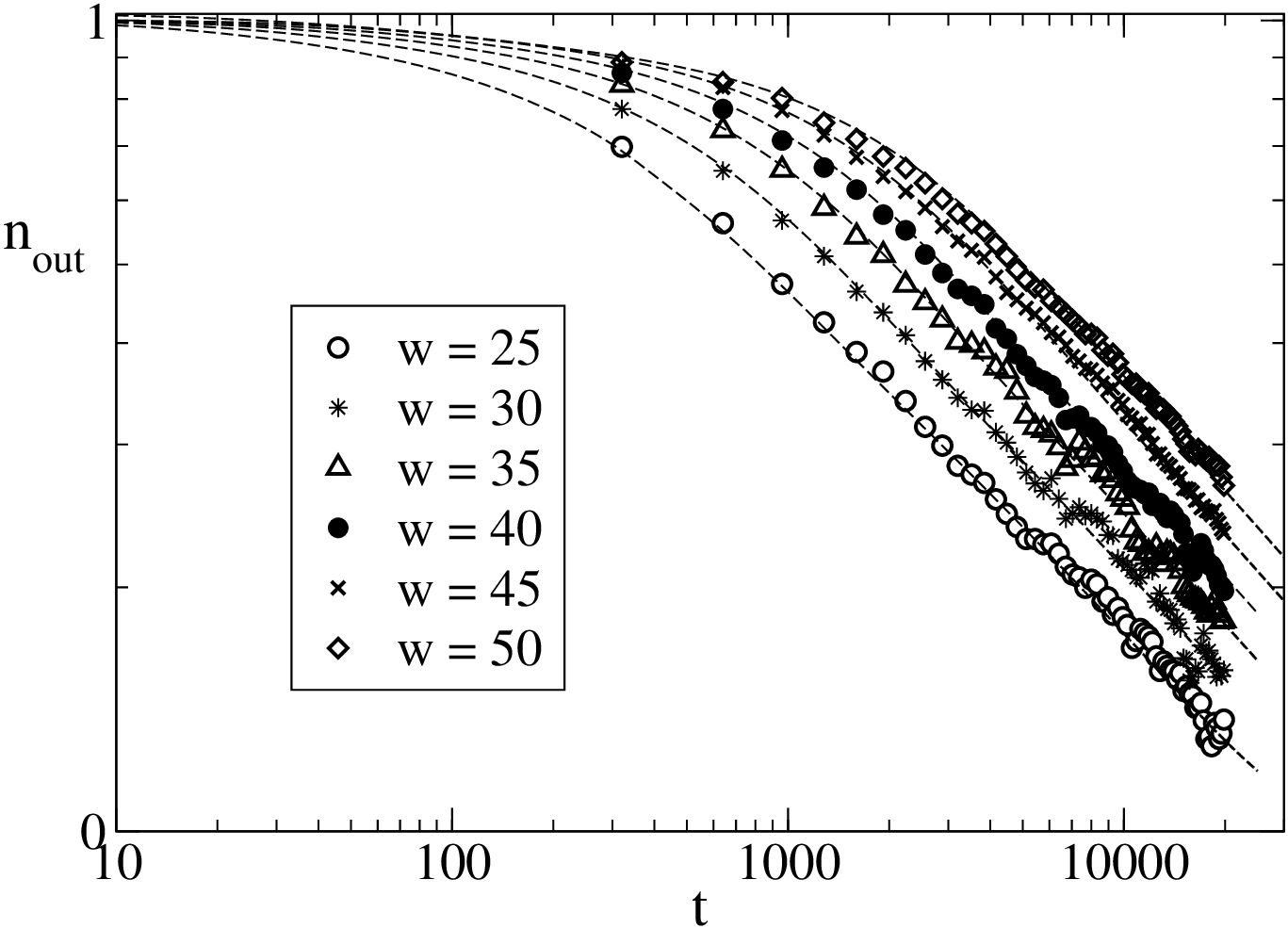}
\caption{Time decay of the number of particles 
initially located in a single well of width $w$
for $\alpha=0.8$, $\Gamma=5$. The same numerical experiment is repeated
for different well sizes but maintaining fixed the  
barrier height. Broken lines represent the theoretical predictions
and the symbols the simulations results obtained as an average over 
$300$ independent runs.}
\label{fig:escape}
\end{figure}

Finally, we consider the escape process of a group of particles 
from a well of the sinusoidal potential. This numerical experiment
amounts to preparing initially a set of particles in a potential well 
and measuring the fraction $n_{in}(t)$ that populates the 
well at that instant. Figure~\ref{fig:escape} reports the log-log plot 
of decay in time of $n_{in}(t)$, 
obtained by averaging over $300$ independent 
runs, for different well widths $w$. 
For comparison we also show the 
corresponding quantity computed through the DDFT (dashed lines) 
which correctly 
reproduces the dynamical features of the escape process.    
We see that the rate at which particles exit the original well decreases
with $w$, because the particles spend more time in that well.    
The collisionless particle systems generally displays a longer escape time 
than interacting systems, because the absence of excluded volume effects 
does not entail an  
effective reduction of the depth of the potential well, which, instead, is 
the relevant feature characterizing the escape experiments involving hard 
core particles. 
The inelasticity, on the other hand, tends to decrease the escape rate 
due to the energy loss caused by collisions, but this does not 
counterbalance the excluded volume effect so that the rate of the inelastic
particles remains faster than the corresponding rate of the collisionless 
model.

\section{Conclusions} \label{Conclusions}
In this paper we have derived a method to study the dynamics
of an assembly of particles interacting inelastically, 
and driven by a stochastic thermostat.
We have found that the particles adopt 
spatial configurations which are
very close to those of an equilibrium system, in spite of the fact that
our system is driven and dissipative. The reason for such a similarity
is twofold as suggested by a recent study of Reis et al.~\cite{Reis}:
the homogeneous energy feeding mechanism and the importance of the
repulsive forces.  However, the present
work shows that there is no need to invoke entropic forces
to explain the observed inhomogeneities. 
A kinetic approach, in which
the short range repulsion is suitably
accounted for by means of a suitable treatment of spatial correlations,
predicts fairly well the observed structural properties~\cite{Urbach}.
Our theory indicates that the steady state configurations occur 
not as a result of the minimization of some hypothetical coarse grained
free energy functional but as a result of the competition between the
uniform energy injection and the energy dissipation.
These two effects are described by an Enskog collision operator and
by a a  Fokker-Planck collision operator, respectively.
The evolution of the phase space distribution function is thus governed by
a Fokker-Planck-Enskog (FPE) equation, whose solution still
remains an extremely difficult
task for dense fluids due to the complexity of the collision kernel
and to the computer resources needed to resolve the distribution function.
However, when the friction is sufficiently high one can derive
a simpler description
by an iterative elimination of the fast 
degrees of freedom,
such as the velocities of the particles. Such a procedure 
is based on the intuition
that these achieve locally their equilibrium distribution, whereas
the positions evolve more slowly. 
Truncating this iteration at
the first order in the inverse friction parameter
$\Gamma^{-1}$ is equivalent to approximating the velocity
distribution functions by Maxwellians at temperatures equal to that
of the heat bath. Further terms, associated with non Maxwellian
contributions to the velocity distribution function, 
are included in the expansion and contribute to the evolution.
The result is a self-consistent time dependent equation for the local density,
where the ``internal field''  is determined by the density itself either
through the standard hard-rod entropic contribution
or by the velocity and energy currents generated by spatial density gradients.
 
At a technical level Eq.~(\ref{psi0timeb}) is derived  
by applying a multiple time-scale method
to the Fokker-Planck-Enskog equation. The resulting equation bears a strong
similarity with the Dynamic Density Functional equation, but it is
not based on the notion of coarse grained Free Energy, a concept which cannot
be applied to open non equilibrium systems, such as the inelastic fluid we
have studied in this work. 
Equation~(\ref{psi0timeb}) is a density functional equation for the density field
$\phi_{00}(X,\tau)$ where its evolution depends on a functional
of  $\phi_{00}$ itself and its derivatives. It is local in time, but
does not possess a generating Liapunov functional, therefore, 
we cannot prove that the associated dynamics minimizes some
cost function. 

In more detail we have found that: 

a) the inelasticity induces changes even in the
stationary properties of the fluid with respect to the elastic reference
system to order $\Gamma^{-1}$. The changes can be observed 
both in the velocity  distribution and in the structure of the non uniform
fluid. 

b) Whereas in the description of a colloidal fluid the friction
$\Gamma$ is relevant only for the relaxation properties,
in the inelastic fluid $\Gamma$ determines its stationary properties also.

c) The theory holds in the region $\omega_E/\gamma<1$, 
when the typical time-scale
of the heat bath is shorter than the Enskog collision frequency.

As far as future perspectives are concerned the method can be generalized 
to higher dimensions and different types of inter-particle forces
and to systems with a non uniform distribution of heat sources 
\cite{Lopez}.
A second type of generalization consists in performing the same
multiscale expansion at the level of the two particle phase-space
distribution function, by truncating the BBGKY hierarchy one step further, 
which would allow us to compute self-consistently
the pair correlation function of the system.

\section{Acknowledgments}
UMBM acknowledges a grant COFIN-MIUR 2005, 2005027808.
PT acknowledges grants FIS2004-05035-C03-02 by the Direccion General 
de Investigacion
of Spain, and S-0505/ESP/0299 by the Comunidad Autonoma de Madrid.

\appendix 
\section{Collision integrals}

In this appendix we show how to perform the velocity integrations
and reduce the collision integrals to simple functions
of space and time only.
Using the definition of collision integral
given in the text by Eq.~(\ref{ret2}) and setting
$u=V_2-V$, we obtain the following explicit expression:
\begin{eqnarray}
\label{a1bis}
C_{n}(X,\tau)&=&g_2(X,X+1)\Bigl\{
\int_{-\infty}^{\infty}dV \mu_{n}(V)
\Bigl[\int_{-\infty}^{0}du u 
P(X,V,\tau)P(X+1,u+V,\tau)\nonumber\\
&+&\frac{1}{\alpha^2}\int_0^{\infty} du u 
P(X,V+su,\tau)P(X+1,V+qu,\tau)\Bigl]\Bigl \}
\nonumber\\
&-&g_2(X,X-1)\Bigl\{
\int_{-\infty}^{\infty}dV \mu_{n}(V)\frac{1}{\alpha^2}
\Bigl[\int_{-\infty}^0 du u 
P(X,V+su,\tau)P(X-1,V+qu,\tau)\nonumber\\
&+&\int_{0}^{\infty}du u 
P(X,V,\tau)P(X-1,u+V,\tau)\Bigl]\Bigl \}\;,
\end{eqnarray}
where $s=(1+\alpha)/2\alpha$, $q=-(1-\alpha)/2\alpha$, 
$\mu_0(V)=1$, $\mu_1(V)=V$ and $\mu_2(V)=V^2/2$. 
After substituting the expression of $P(X,V,\tau)$ in terms of its
partial amplitudes into Eq.~(\ref{a1bis}), one can eliminate the
velocities obtaining:
\begin{eqnarray}
C_{n}(X,\tau)&=&g_2(X,X+1)\sum_{\mu,\nu}
\bigg[N_{\mu\nu}^{(n)}(\alpha=1)+\frac{1}{\alpha^2}
M_{\nu\mu}^{(n)}(\alpha)\bigg]
\Phi_{\mu}(X,\tau)\Phi_{\nu}(X+1,\tau)
\nonumber\\
&-&g_2(X,X-1)\sum_{\mu,\nu}
\bigg[M_{\mu\nu}^{(n)}(\alpha=1)+\frac{1}{\alpha^2} 
N_{\nu\mu}^{(n)}(\alpha)\bigg]
\Phi_{\mu}(X,\tau)\Phi_{\nu}(X-1,\tau).
\label{coll_a}
\end{eqnarray}
where the matrix elements $M_{\mu\nu}^{(n)}$
and $N_{\mu\nu}^{(n)}$ are defined as:
\begin{equation}
N_{\mu\nu}^{(n)}(\alpha)=\int_{-\infty}^0 du u 
\int_{-\infty}^{\infty}dV \mu_{n}(V) 
H_{\mu}(V+qu)H_{\nu}(V+su)
\end{equation}
\begin{equation}
M_{\mu\nu}^{(n)}(\alpha)=
\int_{0}^{\infty}du u\int_{-\infty}^{\infty}dV \mu_{n}(V) 
H_{\mu}(V+qu)H_{\nu}(V+su).
\label{gmunu2}
\end{equation}
and have the symmetry property
\begin{equation}
N_{\mu\nu}^{(n)}(\alpha)=(-1)^{\mu+\nu+n+1}M_{\mu\nu}^{(n)}(\alpha)\;,
\label{symp}
\end{equation} 
so that it is sufficient to calculate only the matrix elements of
$M_{\mu\nu}^{(n)}(\alpha)$ in order to compute Eq.~(\ref{coll_a}):

\[ \frac{M_{\mu\nu}^{(1)}(\alpha)}{\alpha^2}     = \left| 
\begin{array}{ccc}
-\alpha/2 & \frac{1-2\alpha}{2\sqrt \pi} & \frac{1}{2}(1-\alpha)  \\
\frac{1}{2\sqrt \pi}(1+2\alpha) &\alpha/2  &-\frac{1}{4\sqrt{\pi}}(1-2\alpha)\\
-\frac{1}{2}(1+\alpha) & -\frac{1}{4\sqrt{\pi}}(1+2\alpha) & 0 \\
\label{pab1}
\end{array}
\right|.\] 

\[ \frac{M_{\mu\nu}^{(2)}(\alpha)}{\alpha^2}     = \left| 
\begin{array}{ccc}
-\frac{1}{4\sqrt{\pi}}(1-2\alpha^2) & -\frac{1}{8}(1+2\alpha-3\alpha^2) & 
\frac{1}{8\sqrt{\pi}}(1-8\alpha+6\alpha^2) \\
\frac{1}{8}(1-2\alpha-3\alpha^2) & \frac{3}{8\sqrt{\pi}}(1-2\alpha^2) 
&\frac{1}{8}(1-\alpha)(1+3\alpha)\\
\frac{1}{8\sqrt{\pi}}(1+8\alpha+6\alpha^2)&-\frac{1}{8}(1+\alpha)(1-3\alpha)& 
-\frac{3}{16\sqrt{\pi}}(1-2\alpha^2)\\
\label{pab2}
\end{array}
\right|.\] 

In addition, we verify that
in the case $n=0$ the following combinations vanish:
$$
N_{\mu\nu}^{(0)}(\alpha=1)+\frac{1}{\alpha^2}
M_{\nu,\mu}^{(0)}(\alpha)=0
$$
$$
M_{\mu\nu}^{(0)}(\alpha=1)+\frac{1}{\alpha^2}
N_{\nu\mu}^{(0)}(\alpha)=0
$$ 
so that $C_0(X,\tau)=0$, since collisions conserve the number of particles.

\section{Multiple time scale method.}
Our previous work~\cite{Tarazona2} has extended 
to the case of colliding particles
a method to derive the Smoluchowski equation
starting from the Kramers equation. It was originally proposed 
in the 1970s for a gas of non-interacting particles by
Titulaer~\cite{Titulaer} and nicely reviewed by 
Bocquet~\cite{Bocquet,Hansen}.

It represents  a particular application of multiple time-scale 
analysis~\cite{Bender} designed to handle singular perturbations.
In the present case the singularity stems from the fact
that when $\Gamma>>1$ the time derivative occurs among the small terms
of Eq.~(\ref{brinkmanhiera}). 

Because the inelasticity brings about some
remarkable new features we shall report the derivation of the salient
parts of the multiple scale method in this particular case. 
The multiple time-scale analysis introduces a set of auxiliary
time scales $\tau_n=\Gamma^{-n}\tau$,  with $n=0,1,2,...$.
The $\tau_n$ are treated as independent variables so that the time derivative
with respect to $\tau$ is replaced by
\begin{equation}
\frac{\partial}{\partial \tau}=\frac{\partial}{\partial \tau_0}
+\frac{1}{\Gamma} \frac{\partial}{\partial \tau_1}
+\frac{1}{\Gamma^2} \frac{\partial}{\partial \tau_2} + \cdot
\label{mult}
\end{equation}
The partial amplitudes $\Phi_{\nu}(X,\tau)$
and the collision terms $C_{\nu}(X,\tau)$ are also treated as functions
of the auxiliary time scales and expanded perturbatively as: 
\begin{equation}
\Phi_\nu(X,\tau_0,\tau_1,\tau_2,..)=\sum_{n=0}^{\infty} \frac{1}{\Gamma^n} 
\phi_{n \nu}(X,\tau_0,\tau_1,\tau_2,..)
\label{phi}
\end{equation}
and 
\begin{equation}
C_\nu(X,\tau_0,\tau_1,\tau_2,..)=\sum_{n=0}^{\infty} \frac{1}{\Gamma^n} 
c_{n\nu}(X,\tau_0,\tau_1,\tau_2,..) \qquad .
\label{cexp}
\end{equation}
By substituting Eqs.~(\ref{mult})-(\ref{cexp}) into Eq.~(\ref{brinkmanhiera})
and equating equal powers of $\Gamma$, 
one obtains iteratively a series of equations which must be satisfied by
the coefficients $\phi_{s\nu}(X,\tau_0,\tau_1,\tau_2,..)$ 
and $c_{s\nu}(X,\tau_0,\tau_1,\tau_2,..)$. 
The latter coefficients are obtained using the formula:
\begin{eqnarray}
&&c_{s\nu}(X,\tau)=\sum_{l+m=s}\sum_{\mu,\nu}g_2(X,X+1)
\bigg[N_{\mu\nu}^{(n)}(\alpha=1)+\frac{1}{\alpha^2}
M_{\nu\mu}^{(n)}(\alpha)\bigg]
\phi_{l\mu}(X,\tau)\phi_{m\nu}(X+1,\tau)\nonumber\\
&-&g_2(X,X-1)
\bigg[M_{\mu,\nu}^{(n)}(\alpha=1)+\frac{1}{\alpha^2} 
N_{\nu\mu}^{(n)}(\alpha)\bigg]
\phi_{l\mu}(X,\tau)\phi_{m\nu}(X-1,\tau).
\label{colfin}
\end{eqnarray}
Notice that the $c_{s\nu}$'s are functionals of the $\phi_{s\nu}$'s.

We begin with the order $\Gamma^0$:
\begin{equation}
L_{FP} \Bigr[\sum_\nu \phi_{0\nu}H_\nu\Bigr]=0
\label{g0}
\end{equation} 
having the solution $\phi_{0\nu}=0$ for $\nu\neq 0$,
which inserted in Eq.~(\ref{kramers0}) determines
the expansion coefficients of order $\Gamma^{-1}$ in 
terms of $\phi_{0\nu}=0$.
\begin{eqnarray}
L_{FP} &\Bigr[& \phi_{11}H_1+\phi_{12}H_2 +\phi_{13}H_3 
+\phi_{14}H_4+..\Bigr]=\\\nonumber
&&\frac{\partial \phi_{00}}{\partial \tau_0}H_0+D_X\phi_{00}H_1
-c_{01}H_1-c_{02}H_2-c_{03}H_3-c_{04}H_4..
\label{g1}
\end{eqnarray} 
where we have employed the abbreviation $D_X\equiv(\partial_X-F(X))$.
We also perform our expansion by setting $\phi_{s0}=0$ for all $s>0$.
By equating the coefficients of the same $H_{\nu}$ in Eq.~(\ref{g1})
we find  the following
relations
\begin{equation}
\frac{\partial \phi_{00}}{\partial \tau_0}=0
\label{psi0t}
\end{equation}
\begin{equation}
\phi_{11}=-D_X\phi_{00}+c_{01}
\label{ps11}
\end{equation}
and for $\nu>1$
\begin{equation}
\phi_{1\nu}=\frac{1}{\nu}c_{0\nu}.
\label{ps12}
\end{equation}
The procedure can be iterated to the order $\Gamma^{-2}$, writing
the equation
\begin{eqnarray}
&&L_{FP}\Bigr[\sum_{\nu\geq 1}\phi_{2\nu}H_\nu\Bigr]=
\sum_{\nu\geq 1} \frac{\partial \phi_{1\nu}}{\partial \tau_0}H_\nu
+\frac{\partial \phi_{00}}{\partial \tau_1}H_0
+D_X\phi_{11}H_2+\partial_X\phi_{11}H_0+\\\nonumber
&&D_X\phi_{12}H_3+2\partial_X\phi_{12}H_1
+D_X\phi_{13}H_4+3\partial_X\phi_{13}H_2
+D_X\phi_{14}H_5+4\partial_X\phi_{14}H_3
-\sum_{\nu\geq 1} c_{1\nu}H_\nu
\nonumber
\end{eqnarray}
which leads to the following conditions
\begin{equation}
\frac{\partial \phi_{00}}{\partial \tau_1}=
-\partial_X\phi_{11}
\label{psi0t1}
\end{equation}
\begin{equation}
\frac{\partial \phi_{11}}{\partial \tau_0}=
-\phi_{21}-2\partial_X\phi_{12}+c_{11}=0
\label{psi11t0}
\end{equation}
\begin{equation}
\frac{\partial \phi_{12}}{\partial \tau_0}=
-2\phi_{22}-D_X\phi_{11}- 3 \partial_X\phi_{13} + c_{12}=0,
\label{psi12t0}
\end{equation}
and
\begin{equation}
\frac{\partial \phi_{13}}{\partial \tau_0}=
-3\phi_{23}-D_X\phi_{12}-4\partial_X\phi_{14} + c_{13}=0.
\label{psi13t0}
\end{equation}
The amplitude $\phi_{11}$, being a functional of $\phi_{00}$, 
does not depend on $\tau_0$. Hence, the l.h.s. of Eq.~(\ref{psi11t0})
vanishes and we find
\begin{equation}
\phi_{21}=-2\partial_X\phi_{12}+c_{11}.
\label{psi21b}
\end{equation}
Similarly, we obtain for $\nu>1$:
\begin{equation}
\phi_{2\nu}= -\frac{1}{\nu} 
[D_X\phi_{1(\nu-1)}+(\nu+1) \partial_X\phi_{1(\nu+1)}+c_{1\nu}]
\label{psi2n}
\end{equation}
Explicitly we write
\begin{equation}
\phi_{22}= -\frac{1}{2} 
[D_X D_X\phi_{00}+D_X c_{01} +\partial_X c_{03}-c_{12}]
\label{psi22}
\end{equation}
\begin{equation}
\phi_{23}=-\frac{1}{3}[\frac{1}{2} D_X c_{02}+\partial_X c_{04}-c_{13}]
\label{psi23}
\end{equation}
\begin{equation}
\phi_{24}=-\frac{1}{4}[\frac{1}{3} D_X c_{03}+\partial_X c_{05}-c_{14}].
\label{psi24}
\end{equation}

Finally, by equating the coefficients of $H_0$, we arrive to the equation:
\begin{equation}
\frac{\partial \phi_{00}}{\partial \tau_1}=
\partial_X[D_X\phi_{00}-c_{01}].
\label{psi0t1b}
\end{equation}


In order to carry out the derivative of $\phi_{00}$
with respect to the time $\tau_2$,
we iterate the procedure to the order $\Gamma^{-3}$ by writing: 
\begin{eqnarray}
&&L_{FP}\Bigr[ \sum_{\nu\geq 1}\phi_{3\nu}H_\nu   \Bigr]
=\frac{\partial  }{\partial \tau_0}\sum_{\nu\geq 1}\phi_{2\nu}H_\nu+
\frac{\partial }{\partial \tau_1}\sum_{\nu\geq 1}\phi_{1\nu}H_\nu
+\frac{\partial\phi_{00}}{\partial \tau_2}H_0\\\nonumber
&+&\sum_{\nu\geq 1}[D_X\phi_{2\nu}H_{\nu+1}+
\nu\partial_X\phi_{2\nu}H_{\nu-1}]-\sum_{\nu\geq 1}c_{2\nu}
\nonumber
\end{eqnarray}
and equating  the coefficients of $H_0(V)$ we 
obtain the following equation 
\begin{equation}
\frac{\partial \phi_{00}}{\partial \tau_2}=
-\partial_X\phi_{21}=-\partial_X c_{11}+\partial_X^2 c_{02}
\label{psi0t2}
\end{equation}
where the second equality follows from eqs.(\ref{ps12}) and (\ref{psi21b}).

We, now, collect together the different orders in $\tau_n$, given by 
Eqs.~(\ref{mult}), (\ref{psi0t1}) and (\ref{psi0t2}), thus
restoring the original physical time $\tau$ to find
the evolution equation for the density amplitude (\ref{psi0timeb}).



\begin{thebibliography}{0}
\bibitem{footnote1}
We remark that in the present model two features concur to
remove the pathological aspects of one dimensional hydrodynamics: 
the stochastic driving and the inelasticity of collisions.
\end{thebibliography}
\end{document}